\documentclass[conference]{IEEEtran}
\IEEEoverridecommandlockouts
\usepackage{cite}
\usepackage{amsmath,amssymb,amsfonts}
\usepackage{algorithmic}
\usepackage{graphicx}
\usepackage{textcomp}
\usepackage{xcolor}
\usepackage{tcolorbox}
\usepackage{wrapfig}
\usepackage{amsfonts}
\usepackage{graphicx}
\usepackage{stfloats}
\usepackage{enumitem}
\usepackage{xspace}
\usepackage{comment}
\usepackage{caption}
\usepackage{epstopdf}
\usepackage{algorithmic}
\usepackage{algorithm2e}
\usepackage{tabularx}
\usepackage{setspace}
\usepackage{xspace}
\usepackage{xcolor}
\usepackage{wrapfig}
\usepackage{array}
\usepackage{makecell}
\usepackage{lipsum}
\usepackage{tcolorbox}
\usepackage{todonotes}
\usepackage{tabularx,booktabs}
\newcolumntype{C}{>{\centering\arraybackslash}X} 
\usepackage{caption}

\usepackage{subcaption}
\usepackage{euscript}
\newcolumntype{P}[1]{>{\centering\arraybackslash}p{#1}}
\usepackage[flushleft]{threeparttable}
\usepackage{array,booktabs,makecell}
\usepackage{booktabs}
\usepackage{dcolumn,tipa}
\newcolumntype{d}[1]{D{.}{.}{#1}} 
\usepackage[usestackEOL]{stackengine}
\usepackage{multirow}

\usepackage{tabulary}
\usepackage[normalem]{ulem} 
\usepackage{multirow,array}
\usepackage{hyperref}   
\usepackage{makecell}
\usepackage{pdflscape}  
\usepackage{colortbl}
\usepackage{soul}
\usepackage{algorithm2e}
\RestyleAlgo{ruled}
\RestyleAlgo{vlined}
\SetKwInput{KwInput}{Input}
\SetKwInput{KwOutputput}{Output}

\usepackage{makecell}

\definecolor{light-gray}{gray}{0.96}
\definecolor{LightCyan}{rgb}{0.88,1,1}

\definecolor{light-gray}{gray}{0.96}

\usepackage{amsmath}
\DeclareMathOperator*{\argmax}{arg\,max}

\newcommand{\ipAddress}[1]{{\fontfamily{pcr}\selectfont #1}}

\newif\ifcomments
\commentstrue

\newcommand{\AlgoName}{\textsc{CAPoW}\xspace}

\newcommand{\ModelA}{DAbR\xspace}
\newcommand{\ModelB}{\textsc{TAM}\xspace}
\newcommand{\ModelC}{\textsc{Flow}\xspace}

\newcommand{\ScoreA}{{\alpha}\xspace}
\newcommand{\ScoreB}{{\beta}\xspace}
\newcommand{\ScoreC}{{\gamma}\xspace}

\newcommand{\Score}{\textsc{$\Phi$}\xspace}

\begin{document}

\title{CAPoW: Context-Aware AI-Assisted Proof of Work based DDoS Defense}
\author{\IEEEauthorblockN{Trisha Chakraborty\IEEEauthorrefmark{1},
Shaswata Mitra\IEEEauthorrefmark{2}, Sudip Mittal\IEEEauthorrefmark{3}}
\IEEEauthorblockA{Department of Computer Science \& Engineering, Mississippi State University\\
\{{tc2006\IEEEauthorrefmark{1},
sm3843\IEEEauthorrefmark{2}\}@msstate.edu},
mittal\IEEEauthorrefmark{3}@cse.msstate.edu}}
\maketitle
\pagestyle{plain} 
\begin{abstract}

Critical servers can be secured against distributed denial of service (DDoS) attacks using proof of work (PoW) systems assisted by an Artificial Intelligence (AI) that learns \textit{contextual} network request patterns. In this work, we introduce \AlgoName, a \textit{context-aware} anti-DDoS framework that injects latency \textit{adaptively} during communication by utilizing context-aware PoW puzzles. In \AlgoName, a security professional can define relevant request context attributes which can be learned by the AI system. These contextual attributes can include information about the user request, such as IP address, time, flow-level information, etc., and are utilized to generate a contextual score for incoming requests that influence the hardness of a PoW puzzle. These puzzles need to be solved by a user before the server begins to process their request. Solving puzzles slow down the volume of incoming adversarial requests. Additionally, the framework compels the adversary to incur a cost per request, hence making it expensive for an adversary to prolong a DDoS attack. We include the theoretical foundations of the \AlgoName framework along with a description of its implementation and evaluation. 
\end{abstract}



\section{Introduction}
\label{sec:introduction}

An organization protects its critical servers from distributed denial of service (DDoS), which may contain valuable information, such as intellectual property, trade secrets, employee personally identifiable information (PII), etc. To launch a DDoS attack, the malicious users send a flood of requests to these servers. As a result, requests from legitimate users either experience delays or their requests are dropped. For more than two decades, DDoS attacks have been a prominent issue and even today it is far from being solved as these attacks are cheaper to launch than to defend, especially with the rise of DoS-as-a-Service~\cite{DDoS-Saas}.

PoW system works by requiring incoming requests to expend resources solving an \textit{computational puzzles} to prove ones legitimacy. The general system consists of two parts: \textit{prover} and \textit{verifier}. The prover finds the solution to the computational puzzles, when solved, sends the solution to the verifier. In a simple networked client-server environment, the user-side contains the prover component, and the server-side contains the verifier components. Researchers have proposed PoW-based solutions for DDoS which makes the attack expensive to launch~\cite{AURA2001DOS, waters:new, mankins2001mitigating}.  Although, these solutions suffer from a lack of intuition on how to set puzzle difficulty and adaptability in different settings. 

In this paper, we develop a defensive tool that emphasizes on learning the normal activity patterns of legitimate users. The idea behind the tool is to penalize the users that \textit{deviates} from normal activity patterns by issuing them \textit{hard} puzzles and at the same time issuing \textit{easy} puzzles to users who follow the pattern. We leverage a \textit{context-aware AI model} that can learn these normal activity patterns by contextual information. The term \textit{context} within the scope of legitimate activity patterns can be defined as request attributes, such as, IP address, time, flow-level information, etc. When the context is \textit{IP address}, network activity is considered deviated if the source IP address is part of a known blocked IP list. Whereas, when the context is \textit{time}, network activity is considered deviated if it arrives at an unusual time compared to the normal activity pattern. Security professionals can select relevant request context attributes which can be learned by the AI models. The concept of \textit{context-aware AI models} is derived from context-aware computing introduced by Dey et. al~\cite{Anind2001Context}.

We introduce \AlgoName tool, a \textit{context-aware} AI-assisted PoW system that helps to secure critical servers against DDoS attacks. Our framework utilizes context-aware AI models that learn the expected context pattern from server-side activity-logs. The activity-logs are stored and managed by the server which contains user activity (IP address, timestamp, flow-level data, etc).  The deviation from the learned pattern is then leveraged to generate a \textit{contextual score} for incoming requests which tunes the difficulty level of the PoW puzzle to be solved.  \textit{The underlying defensive strategy curtails the ability of a malicious user to prolong the attack by \textit{adaptively} introducing \textit{latency} through PoW puzzles and compelling malicious users to expend more resources to complete an attack.}  The main contributions of this paper are as follows.

\noindent\textbf{Contribution 1:} We introduce \AlgoName, an anti-DDoS framework that injects
latency adaptively, i.e., the framework ensures that malicious users incur higher latency than legitimate users based on the deviation in context pattern. We discuss the process of context score calculation from deviation in Section~\ref{sec:ai-theory}.

\noindent\textbf{Contribution 2:} We propose a policy component that is created by security personnel to incorporate server-specific security demands. We provide intuition for policy construction in Section~\ref{sec:policy-theory}.

\noindent\textbf{Contribution 3:} We discuss an instance of \AlgoName implementation and perform evaluation to illustrate the effectiveness of \AlgoName. The implementation details are discussed Section~\ref{sec:evaluation}. The code is released on GitHuB~\cite{OnlineCode}. 

The rest of the paper is structured as follows. In Section~\ref{sec:threat-model} we discuss the threat model and attack definitions. We discuss the theoretical foundation of \AlgoName in Section~\ref{sec:tooldesign} and \AlgoName implementation in Section~\ref{sec:evaluation}. We discuss related works of the PoW system and DoS defense in Section~\ref{sec:related-works}, followed by the conclusion in Section~\ref{sec:conclusion}.


\section{Threat Model}
\label{sec:threat-model}
In this section, we present a series of assumptions associated with the adversary's abilities. An adversary $\mathbb{A}$ initiates a DDoS attack by sending a flood of requests to the server. The adversary's intention is to overwhelm the server's computational resources and disrupt legitimate user communication with the server. Although the attack described is a variant of DDoS, the usefulness of \AlgoName can be extended to other variants. These assumptions described below are similar to previous literature on DDoS defense using proof of work~\cite{juels:dos} and in some sense, we consider a stronger adversary.\vspace{2mm}

\noindent\textbf{Assumption 1}. Adversary $\mathbb{A}$ can eavesdrop on the communication channel of the server. $\mathbb{A}$ cannot modify any user request and cannot read any request payload data.\medskip

Assume a secure network communication channel is used by the user to send request packets to the server. The user performs encryption on the payload data, including the puzzle solution, and sends the packet to the server. When an adversary eavesdrops on the channel, they can read the source and destination IP of the packet, but they cannot read the encrypted payload consisting of the puzzle parameters. Additionally, the adversary cannot flip bits of the packet and pollute the puzzle solution included in the payload. Hence, we assume that the adversary has no knowledge of the puzzle parameters solved by a user nor can it deny service to a user who has correctly solved the puzzle. In Section~\ref{sec:evaluation}, we utilize assumption 1 to claim that the adversary cannot reverse engineer the base AI models to receive easier PoW puzzles.\vspace{2mm}

\noindent\textbf{Assumption 2} Adversary $\mathbb{A}$ can spoof user identifiers, such as IP addresses, and deceive a subset of underlying AI models.\vspace{2mm}

\AlgoName uses AI models to learn legitimate network activity patterns and the deviation from the pattern is directly proportional to the difficulty of PoW puzzles to be solved by the user. $\mathbb{A}$ can spoof a legitimate user IP address and send requests to the server. An intelligent adversary would send probe packets to the server using a set of spoofed IP addresses and only utilize IPs that require puzzles to be solved. This way, the adversary is able to deceive the AI model and reduce the latency introduced. In Section \ref{sec:evaluation}, we discuss that sending probe packets becomes costly for an adversary to deceive multiple base AI models. \vspace{2mm}

\noindent\textbf{Assumption 3} Adversary $\mathbb{A}$ cannot pollute the training data of the AI models.\vspace{2mm}

The AI model used by \AlgoName learns normal activity patterns and calculates a deviation which directly influences the hardness of the puzzle. Hence, it is essential that the AI learns normal activity patterns from an unpolluted activity-log to maximize the effectiveness of \AlgoName. In Section~\ref{sec:ai-implementation}, we describe the training process of a context-aware AI model where a security professional is deployed to select secure data to train the base AI models.


\section{\AlgoName Architectural Design and Theoretical Foundations}
\label{sec:tooldesign}

\begin{figure*}[]
  \centering
  \includegraphics[width=\textwidth, trim = 7cm 9cm 12cm 10cm, clip]{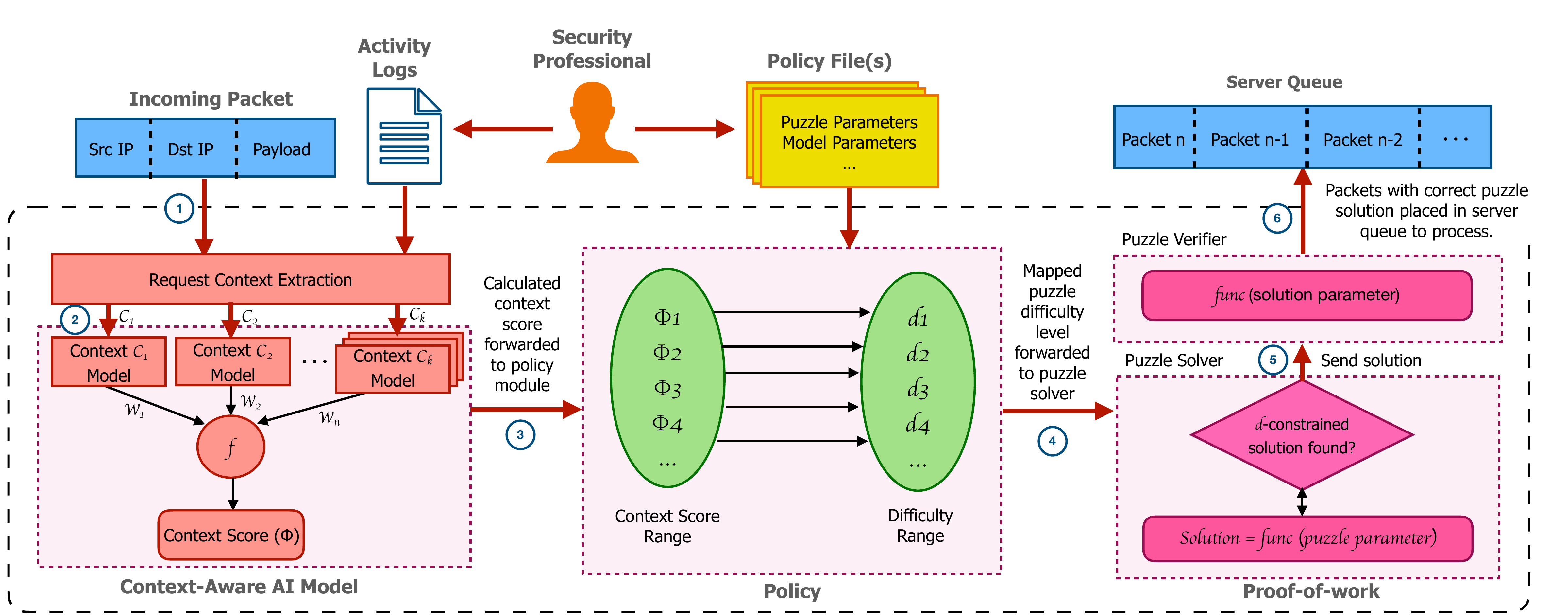}
  \caption{The figure illustrates the architecture of \AlgoName framework. \AlgoName consists of four core components: request context extractor, context-aware AI model, policy, and proof of work. The AI model learns context patterns from previous activity-logs selected by security personnel and calculates a context score based on the deviation of the incoming packet. The calculated score is mapped to the PoW puzzle difficulty level as defined by the security professional in policy files. The proof of work component performs evaluations to find the constrained solution. The request with a correct solution is placed on the server queue to process.} 
  \label{fig:architecture}
  \vspace{-3mm}
\end{figure*}


In this section, we describe the high-level architecture of the core components and their inner workings that molds the \AlgoName framework. As shown in Figure~\ref{fig:architecture}, \AlgoName consists of four core components: \textit{request context extractor, context-aware AI models, policy}, and \textit{proof-of-work}.

The AI models learn the normal activity pattern from previous activity-logs. When an incoming request packet is seen, first the context attributes are extracted from the new request packet (see Section~\ref{sec:context-theory}). Then, the deviation between the learned normal context pattern and new request contexts is computed to calculate \textit{context score}.  We elaborate on AI model training and score calculation in Section~\ref{sec:ai-theory}. The policy component of \AlgoName provides security professionals with certain abilities that strengthen the effectiveness of \AlgoName in various security settings (see Section~\ref{sec:policy-theory}). The context score influences the difficulty of PoW puzzle. In Section~\ref{sec: pow-theory}, we discuss the proof-of-work component and how the PoW puzzles can curtails the ability of a malicious user to prolong the attack by adaptively introducing latency. \vspace{1mm}

\noindent\textbf{Data Flow.} From Figure~\ref{fig:architecture}, the flow of data between different components of \AlgoName is described below. (1) When a new incoming packet is seen, the request packet is forwarded to the request context extractor. (2)  The extracted request context attributes are passed to context-aware AI models which learned expected context patterns from activity logs. The context score generated by individual AI models is combined using a function $f$ to produce the final context score ($\Phi$). (3) The context score is forwarded to the policy component which sets certain parameters, such as, it maps the context score to a puzzle difficulty level. (4) The difficulty level is passed to the puzzle solver which solves a puzzle of the defined difficulty level using a function \textit{func}. (5) The computed solution is sent to the verifier. (6) When the solution is correct, the request packet is placed on the server queue for processing.


\subsection{Context Extraction from Request Packet}
\label{sec:context-theory}
The concept of context-aware computing was introduced by Dey et. al~\cite{Anind2001Context}, where the proposed mechanism improved human-to-computer interaction by delivering contextually relevant data. In the paper, the author proposed an abstract definition of \textit{context}, which is a piece of information that clarifies the characteristics of an entity. When a system contains contextual data about a situation or entity, the system can take context-aware decisions which improve the overall quality of any general decision-making. 

In a security setting, a certain request is deemed suspicious if the associated request attributes deviate from the usual network activity pattern. For instance, a request packet of payload size $65500$ bytes is considered suspicious due to deviation when the expected normal payload size pattern is in the order of a few hundred bytes. To this end, we define \textit{context} of a request packet as request attributes, such as source IP address, time of arrival, port address, time to live (TTL), and other flow-level attributes. The contexts attributes to be extracted are selected by security personnel via policy component. The list of selected context attributes are reformed periodically to update the defensive posture of the organization deployed. When a new request packet is seen, the request context extractor component extracts the selected context attributes from the request packet and feeds it to the context-aware AI models. \vspace{-3mm}


\subsection{Context-Aware AI Model}
\label{sec:ai-theory}
The framework component consumes activity-logs supplied by security personnel as input to generate a context-aware AI model. The model is generated by considering a set of request packets from the activity-log $\lambda=\{\lambda_0,\lambda_{1},\lambda_2,...,\lambda_i\}$. Each request packet $\lambda_i$ consists of a set of request context attributes,
\begin{equation}
    \mathbb{C}_{\lambda_i} = \{\mathbb{C}_{0\lambda_i}, \mathbb{C}_{1\lambda_i}, \mathbb{C}_{2\lambda_i}, ..., \mathbb{C}_{k\lambda_i}\}
\end{equation}
where $k$ is the number of request context attributes. $\mathbb{C}_k$ is represented as $n$-dimensional vector. When an $n$-dimensional vector of a single context for $\lambda$ requests is projected in Euclidean space, such relative positioning produces a cluster. For $k$ context attributes, $k$ clusters are generated. The clusters represent the normal activity pattern. To evaluate a new incoming request, request context extractor from Section~\ref{sec:context-theory}, feeds the context attributes which are then projected in Euclidean space. The deviation $\Delta(p,q)$ of context $\mathbb{C}_k$ is calculated as the Euclidean distance between the corresponding normal activity cluster and the new request projection,
\begin{equation}
   \Delta(p,q) =\sqrt{ \sum_{j=1}^n (q_j - p_j)^2 }
\end{equation}
where $p$ is projected single context attribute of the new request and $q$ is center of a normal cluster of the same context. Consequently, the context score $\Phi$ for $\mathbb{C}_k$ is calculated as,
\begin{equation}
    \Phi(\mathbb{C}_k) = \left(\frac{\Delta(p,q)}{\Delta_{max}}\right)\times I
\end{equation}
where $\Delta_{max}$ is the maximum possible deviation for $\mathbb{C}_k$. The score is in the range of $[0,I]$, where $I\in\mathbb{Z}^+$. In Section~\ref{sec:ai-implementation}, we discuss the implementation of context-aware AI models.


\subsection{Policy}
\label{sec:policy-theory}
The policy component is a rule-based strategy that facilitates the adaptive security guarantees of \AlgoName. The rules are set in policy files that determine certain \AlgoName characteristics. These characteristics include context-aware AI model specifications, such as, which activity-logs are supplied to train the AI models, which context attributes hold more significance over the others, etc. Additionally, these parameters include proof-of-work components specifications, such as, what is the rule to translate context score to puzzle difficulty, which variant of PoW puzzle to be used, etc. Hence, it is evident that policy construction is a non-trivial task and requires consideration of various facets of the deployed server to bolster the effectiveness of \AlgoName in different security settings. To perform the convoluted task of policy designing, \textit{security professionals} are deployed to design server-specific policies.\vspace{1mm}

\noindent\textbf{Intuition for AI model parameters.} From Section~\ref{sec:context-theory}, a request packet consists of several context attributes. The significance of some contexts holds more importance over others depending on the type of attack defense. For instance, payload size is an important context attribute to protect against large payload DDoS attacks~\cite{Ozdel2022Payload}, but less important to defend volumetric DDoS attacks. Policy includes the weight associated with context attributes to provide an attack-specific defense. Additionally, a policy includes the source of data to train the AI models to avoid model data pollution attacks (Assumption 3).\vspace{1mm}

\noindent\textbf{Intuition for proof-of-work parameters.} The context score produced by the context-aware AI model is translated to the PoW difficulty level. The policy includes the rules to translate context scores to puzzle difficulty. In Section~\ref{sec:policy-implemenataion}, we implemented three rules to show that the translation leads to adaptive latency injected. As stated by Green et. al~\cite{Green2011Reconstructing}, amongst groups of users, the CPU capacity of each device can vary $10$x times, whereas memory capacity may only vary $4$x times. Hence,  when a memory-bound PoW puzzle is used, it is less likely for the adversary to have an edge over a legitimate user as the discrepancy in memory power as the resource is less compared to CPU-bound puzzles. The policy includes the means to set variants of puzzles depending on the expected user base. \vspace{1mm}


\subsection{Proof of Work} 
\label{sec: pow-theory}

Classical proof of work systems~\cite{Dwork1992Pricing,waters:new,AURA2001DOS} consists of two main components -- \textit{prover} and \textit{verifier}. The prover provides verifiable evidence of expanding computational resources by solving puzzles as assigned by the server. On the other hand, the verifier validates whether the solved puzzle yielded the desired solution. When PoW systems are used as DoS defense~\cite{AURA2001DOS,Wood2015DoSE,Parno2007Portcullis}, a user commits some computation resources (CPU cycle, bandwidth, etc.) and \textit{burns} one of these resources for solving the PoW puzzle to prove their legitimacy. 

In \AlgoName, when a user deviates from a normal activity pattern, the PoW component issues a PoW puzzle to request proof of legitimacy. The difficulty level of PoW puzzle is a function of context score. The rule to translate to context score to difficulty level is defined under policy component (Section~\ref{sec:policy-theory}). PoW solver uses a function \textit{func} to solve the assigned difficulty puzzle (see Figure~\ref{fig:architecture}). In general terms, this function injects two types of cost: (1) direct cost of \textit{resource burning}~\cite{gupta2020resource}, and (2) indirect cost of \textit{latency}. The notion of resource burning cost represents the resource consumption of a user, where the resource could be computational power, memory, network bandwidth, or human capital~\cite{gupta2020resource}. This cost directly impacts the ability of the adversary to conduct a DDoS attack as every request requires the adversary to spend real-life resources. The notion of \textit{latency} cost captures the delay in time introduced in communication due to the act of puzzle solving. This cost indirectly impacts the adversarial intent by throttling the rate of adversarial requests reaching the server queue. Both costs ultimately cripple the adversarial capability to prolong an ongoing DDoS attack. 


\section{\AlgoName Implementation, Tool Instance Deployment, and Evaluation}
\label{sec:evaluation}

In this section, we present a deployment of \AlgoName framework by implementing a single instance of each core component: context extractor, context-aware AI models, policy, and proof-of-work. First, the context extractor instance extracts selected request context attributes. Second, the extracted contexts are relayed to context-aware AI model instances where each base AI model is generated using server-side activity-logs. Then, the trained AI models calculate the deviation of selected contexts to produce a context score. Third, we provide three policy designs that maps context score to difficulty of PoW puzzle. Finally, we implemented a hash-based PoW puzzle instance which, over repeated trials, finds the constrained solution of assigned difficulty level. The costs inflicted due to the our puzzle instance are CPU-cycles (resource burning) and time spent (latency). For the purposes of validating our contribution via evaluation, we consider that the main cost injected is latency which, when injected, throttles the rate of adversarial requests.

Now, we will describe our evaluation setup. We split the CIC-IDS2017 dataset~\cite{CICIDS2017} into test and train files where day $1$ to day $5$ (Monday - Thursday) is used to train the models and day $6$ (Friday) is used to evaluate \AlgoName. From day $1$ to day $5$, we deleted the attack traffic to learn normal activity pattern. Consider five users sending requests to the server $\mathcal{U}_1, \mathcal{U}_2, \mathcal{U}_3, \mathcal{U}_4,$ and $\mathcal{U}_5$. We fixed four user identifiers from day $5$ to map the four above-mentioned users. Let the fifth user $\mathcal{U}_5$, be mapped to the user identifier that performs DoS on day $6$. Since, the user identifier in CIC-IDS2017 is IP address, let the mapped IP of user $\mathcal{U}_1, \mathcal{U}_2, \mathcal{U}_3, \mathcal{U}_4,$ and $\mathcal{U}_5$ is represented by \ipAddress{104.20.30.120}, \ipAddress{83.66.160.22}, \ipAddress{37.59.195.0}, \ipAddress{104.16.84.55}, and \ipAddress{205.174.165.73} respectively. Through our evaluation scenario, we provided evidence that \AlgoName injects latency adaptively based on the calculated context score of user $\mathcal{U}_5$ which throttles the adversarial requests and make it expensive for an adversary to prolong a DDoS attack. 


\begin{figure*}[h]
  \centering
  \includegraphics[width=\textwidth, trim = 3cm 15cm 2cm 10cm, clip]{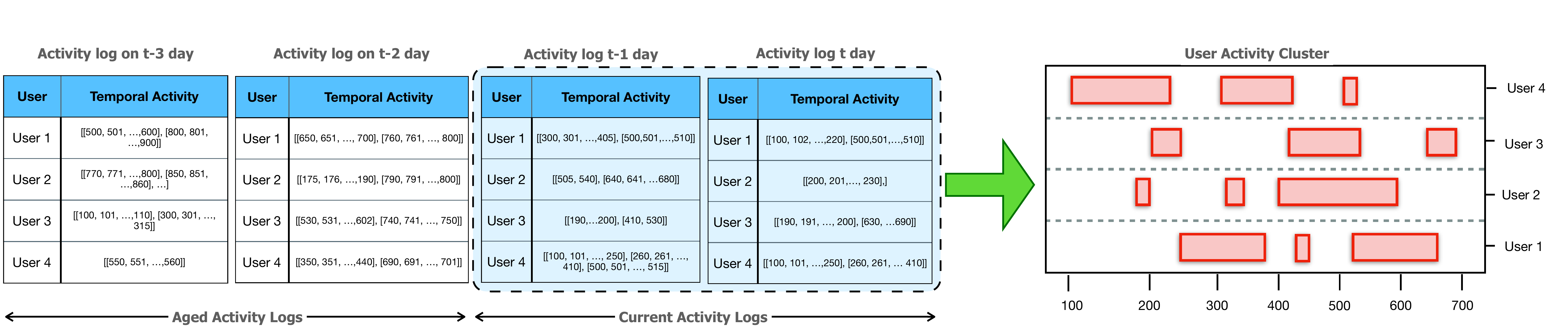}
  \caption[]{The figure shows that selected activity-logs (left) are used to generate a temporal activity model (\ModelB) (right). The illustration shows that out of four activity logs, currently only two activity logs are used to form the model (blue box). The remaining activity-logs are aged in an attempt to keep the model up-to-date.}
  \label{fig:temporal-activity}
  \vspace{-3mm}
\end{figure*}



\subsection{Context Extraction Instance}

The context extraction instance consumes the request packet and extracts context attributes from the request packet. For our implementation, we select three context attributes: (1) IP address, (2) temporal activity, and (3) flow-level data. For evaluation, we used feature attributes of CIC-IDS2017 dataset to serve as context attributes. The source IP feature becomes the IP address context, the timestamp feature becomes the temporal activity context, and the remaining features become flow-level context.


\subsection{Context-Aware AI Model Instance} 
\label{sec:ai-implementation}

We propose an ensemble learner that consists of dedicated base AI models to learn individual contextual patterns. The base AI model receives the context attributes from the context extractor as inputs. The model that (1) learns the IP address pattern is called dynamic attribute-based reputation (\ModelA), (2) learns the temporal activity pattern is called temporal activity model (\ModelB), and (3) learns the flow-level data pattern is called flow-level model (\ModelC). Each model computes a context score in the range between $[0,10]$. Context scores from three AI models are combined using the argmax function. Next, we discuss three base models where the subsections are divided into model generation, context score calculation, and evaluation.\vspace{0.8mm}

\begin{figure*}[h]
  \centering
  \includegraphics[width=\textwidth, trim = 1cm 9.6cm 2cm 13cm, clip]{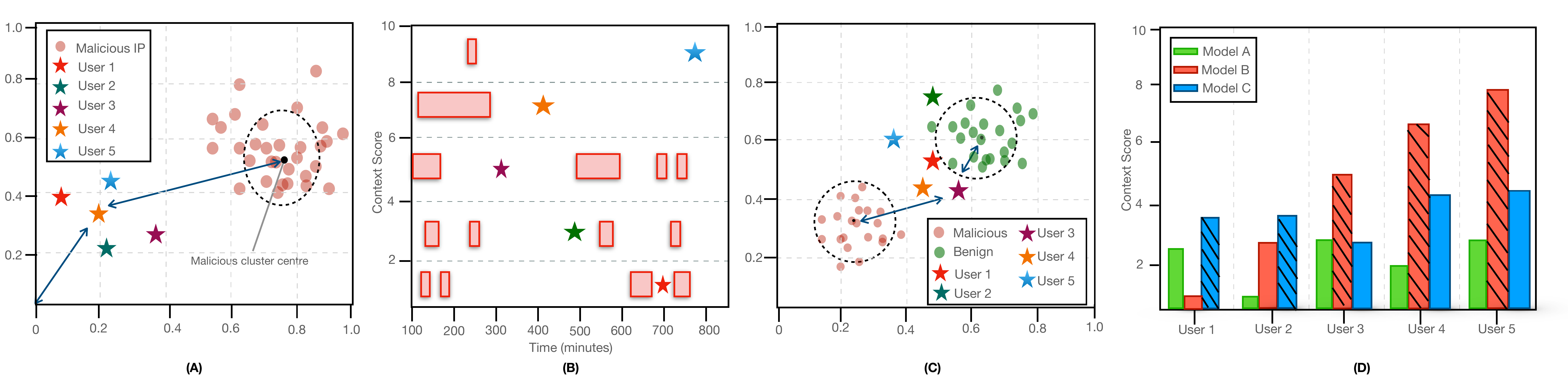}
  \caption[]{The figure contains four sub-figures. (A) Representation of trained \ModelA in the 2-D plot. The red dot cluster represents malicious IP attributes. (B) Representation of trained \ModelB. The stars represent the current time of arrival. (C) Representation of \ModelC. The green cluster represents legitimate flow-level attributes and the red cluster represents malicious ones. (D) Represents the calculated context score after combining scores from Model A is \ModelA, Model B is \ModelB, and Model C is \ModelC.
  }
  \label{fig:evaluation}
  \vspace{-3mm}
\end{figure*}

\noindent\textbf{Dynamic Attribute-based Reputation (DAbR)}: We utilize \ModelA~\cite{renjan2018dabr} as the base AI model that learns context patterns for IP attributes. The AI model is generated by projecting malicious IP attributes from Cisco Talos dataset~\cite{Cisco2022Talos} into Euclidean space. The dataset contains a list of malicious IP addresses and IP-related attributes~\cite{renjan2018dabr}. The red dots in Figure~\ref{fig:evaluation}(A) represent the projected malicious IP attributes that form a cluster in Euclidean space. When a new request is evaluated, the IP attributes of the new request are projected in Euclidean space and a deviation is calculated as Euclidean distance to the malicious cluster center. The distance calculated produces the context score for \ModelA ($\ScoreA$). The multi-colored stars represent $\mathcal{U}_1, \mathcal{U}_2, \mathcal{U}_3, \mathcal{U}_4,$ and $\mathcal{U}_5$. User $\mathcal{U}_1, \mathcal{U}_2, \mathcal{U}_3, \mathcal{U}_4,$ and $ \mathcal{U}_5$ receives $2.87$, $1.16$, $3.15$, $2.18$, and $2.98$ reputation score respectively.\vspace{0.8mm}

\noindent\textbf{Temporal Activity Model (\ModelB)}: We propose a temporal activity model (\ModelB) that learns the pattern of user request activity based on time of arrival from activity-logs. The model is generated using previous $t$-days server activity-logs. The selected activity-logs can be either previous $t$ consecutive days, or $t$ specific days (as defined in the policy). The temporal model can be updated by \textit{aging} the older activity models (see Figure~\ref{fig:temporal-activity}). The red rectangular blocks in Figure~\ref{fig:evaluation}(B) represent an activity cluster per user. The term \textit{active} in practice can represent a user session or concurrent requests. When a user request $\mathcal{U}$ arrives at the server, the server finds the corresponding user activity cluster ($\mathcal{U}_{CLS}$) formed by the temporal activity model. The user activity cluster ($\mathcal{U}_{CLS}$) is a list of time intervals that represents the user's historical activity times. The deviation in time is calculated as the distance between the two nearest clusters. From CIC-IDS2017 dataset, the cluster formed for user $\mathcal{U}_1$ shows that the user was active between $130-140$ minutes, $160-170$ minutes, $600-670$ minutes, and $720-760$ minutes. When user $\mathcal{U}_1$ arrived at time $700$ minutes on day $6$, the two  nearest clusters are $600-670$ and $720-760$ (see Figure~\ref{fig:evaluation}(B)). This deviation is called $\Delta_{local}$ which is the distance between the two nearest clusters. Finally, the context score for \ModelB is calculated as,

\begin{equation}
\label{eq:modelB}
    \ScoreB = \frac{\Delta_{local}}{\Delta_{max}}\times10
\end{equation}

where, $\Delta_{max}$ represents the maximum deviation possible which in our implementation is $720$ minutes. Note that no cluster is found for $\mathcal{U}_5$, hence the context score calculates is the highest in range.
\vspace{0.8mm}

\noindent\textbf{Flow-level Model (\ModelC)}: Flow-level Model (\ModelC) learns network flow context patterns from activity-logs. The network flow attributes of a request packet are flow-related data, such as TTL, flow duration, payload size, protocol, etc. To generate the model, the $n$-dimensional flow attribute vectors are projected in Euclidean space. In Figure~\ref{fig:evaluation}(C), the green dots represent projected network flow attributes of legitimate requests, and the red dots represent projected network flow attributes of malicious requests. When a new request is seen, its flow-level attributes are projected and the Euclidean distance to malicious and legitimate clusters are computed. The context score is calculated as,

\begin{equation}
\label{eq:modelB}
    \ScoreC = \frac{\Delta_{l,m}}{\Delta_{max}}\times10
\end{equation}

where, $\Delta_{l,m}$ is the deviation from malicious and legitimate clusters and $\Delta_{max}$ is the maximum deviation possible in flow-level context.\vspace{1mm}


\subsection{Policy Component Instance}
\label{sec:policy-implemenataion}
We constructed three policy instances, \textit{policy} $1$, \textit{policy} $2$, and \textit{policy} $3$. These policies only set the mapping function between context scores to the PoW puzzle difficulty level. Context score is directly proportional to the difficulty of the PoW puzzle, such as the increase in contextual deviation leads to a higher difficulty puzzle and more latency injected.

\begin{figure}[t]
\centering
 \includegraphics[width=6cm, trim = 1cm 6.5cm 2cm 6cm, clip]{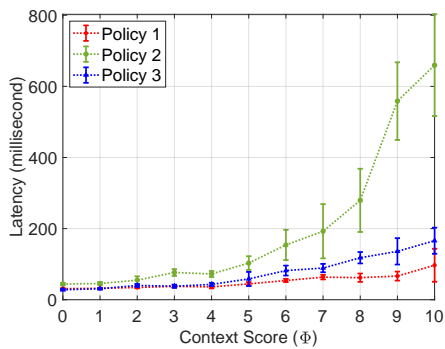}
\caption[]{An evaluation of our three implemented policies. The median of 30 trials is reported for each reputation score.}
\label{fig:policy-plot}
\vspace{-3mm}
\end{figure}

\noindent\textbf{Policies $\mathbf{1}$ and $\mathbf{2}$: Linear mapping}.
Assume a linear map function. Policy $1$ maps $f(\Phi) \rightarrow d $, where $\Phi \in [0,10]$ is the range of context score and $d \in [0,10]$ is the difficulty levels of the PoW puzzle. Similar to policy $1$, policy 2 maps $f(\Phi) \rightarrow d $, where $\Phi \in [0,10]$ and $d \in [10,20]$. Note that, the error bar in Figure~\ref{fig:policy-plot} shows the discrepancy in time to solve $d$-level PoW puzzle. As discussed in Section~\ref{sec:policy-theory}, this discrepancy in time to solve can be avoided by using memory-bound PoW puzzles.\vspace{1mm}

\noindent\textbf{Policy $\mathbf{3}$: Error range mapping}
For policy $3$, we incorporated the error $\epsilon$ of the context-aware AI model. Assume a linear map function. Policy $3$ maps $f(\Phi) \rightarrow d $, where $\Phi \in [0,10]$ and $d \in [0,10]$. The final difficulty level assigned is a difficulty value chosen at random in the interval 
$[\lceil d_i -\epsilon \rceil, \lceil d_i + \epsilon\rceil ]$, where $\epsilon = 0.2$. 
Figure~\ref{fig:policy-plot} shows that as contextual deviation increases, the amount of injected latency increases. 


\subsection{PoW Instance -- Hash Function}
We discuss two sub-components of \AlgoName that mimic proof-of-work system:  \textit{puzzle solver}, and \textit{puzzle verifier}.\medskip

\noindent\textbf{Puzzle Solver.} The puzzle solver takes user identifiers as input, such as the timestamp of the arrival of the request packet ($t$), and the user IP address ($u)$. Additionally, the solver takes a server seed value ($\rho$) to protect against pre-computational attacks. To this, a $n$-bit string is added, which the client modifies upon each hash function evaluation. We call this string \textit{nonce} denoted by $\eta$. 

The user evaluates this input until it finds an output string $Y$ where $Y = H(u||t||\rho||\eta)$ with $d$ leading zeroes, where $d$ is the difficulty level assigned to the request packet. The puzzle solver is a user-end component that is installed either in the browser~\cite{le2012kapow} or kernel-level. After solving, the user sends the nonce back to the server for verification.\vspace{1mm}

\noindent\textbf{Puzzle Verifier.} Puzzle verification is a server-side component that performs straightforward verification of the puzzle solution by performing one hash evaluation, i.e., $Y' = H(u||t||\rho||\eta)$. If the sent $\eta$ value leads to desired number of leading 0's, then the solution is verified. \vspace{1mm}

\noindent\textbf{Summary of \AlgoName implementation evaluation.} 
The context scores produced by \ModelA, \ModelB, and \ModelC models are combined to produce the final context score (\Score). As discussed in Section~\ref{sec:policy-theory}, some contexts might be more relevant than others to provide attack specific defense. We denote weight $w$ as the significance of each context in the final context score. The weights for each AI model are fixed through the policy instance as discussed in Section~\ref{sec:policy-implemenataion}.

\begin{equation}
    \Phi = \argmax(w_{1}\ScoreA, w_{2}\ScoreB, w_{3}\ScoreC)
\end{equation}

where $w_1, w_2,$ and $w_3$ represent weights associated with \ModelA, \ModelB, and \ModelC respectively. Figure~\ref{fig:evaluation}(D) illustrates the combined context score where $w_1, w_2,$ and $w_3$ is set to $1$. User $\mathcal{U}_1$ and $\mathcal{U}_2$ show that the final context score is decided by \ModelC model. Similarly,  $\mathcal{U}_3, \mathcal{U}_4,$ and $\mathcal{U}_5$ the final score is decided by \ModelB model. Using policy $2$, user $\mathcal{U}_5$ incurs $\approx 300$ms latency for a context score of $8$, which is the highest latency amongst other users introduced by \AlgoName.  

Notably, the evaluation performed using a simulated dataset might not reflect the worst case efficiency of \AlgoName as in practice, user $\mathcal{U}_5$ might not be deviate in a temporal activity context. In this section, we discuss that the cost of deceiving multiple AI models is expensive for the adversary. In our implementation, user $\mathcal{U}_5$ has to deceive three AI models to receive an easy PoW puzzle by receiving lower context scores. User $\mathcal{U}_5$ can receive a lower context score for \ModelA by trivially spoofing the IP address (Assumption 2). To deceive \ModelB, the user can engineer the requests around the same time as noticed during eavesdropping (Assumption 1). As reading or tracking flow-level data embedded in request payload data while eavesdropping is not possible (Assumption 1), the only way to deceive \ModelC is by sending multiple probe packets to land on a low context score. This is an extensive approach as a security personnel may select new contexts to improve the defensive posture of the organization periodically. Therefore, deceiving all AI models becomes expensive for the adversary. To validate contribution $3$, we designed and evaluated an implementation instance on \AlgoName and provided policy designs to validate contribution $2$. Finally, \AlgoName ensures that malicious users incur higher latency than legitimate users based on the deviation in context pattern that prevents DDOS. Hence, we validate contribution 1 (see Section~\ref{sec:introduction}). 


\section{Related Works} 
\label{sec:related-works}
In this section, we discuss the overview of proof-of-work (PoW) literature in DDoS. Relevant to our work, we will also discuss the current advances in AI-assisted cybersecurity. 


\subsection{Classical Proof-of-Work}
Dwork et. al~\cite{Dwork1992Pricing}  coined the term proof-of-work (PoW) when they proposed the use of cryptographic hash functions (also known as client puzzles) to combat unsolicited bulk emails (junk emails). Following that, Franklin et. al~\cite{Franklin1997Metering} proposed a lightweight website metering scheme in 1997 to prevent fraudulent web server owners from inflating their website's popularity. In 1999, Jakobsson et. al~\cite{Jakobsson1999Pudding} proposed MicroMinting (originally proposed by Rivest et. al~\cite{Rivest1996PayWordAM} as a digital payment scheme) as a candidate problem that can reuse the computational effort of solving the POW puzzle. Later that year, Laurie et. al~\cite{Laurie2004ProofofWorkPN} proposed that proof of work does not work in a spam setting.


\subsection{Proof-of-Work as DoS defense}

Similar to spam emails, in DDoS, it is significantly cheaper for the attacking party to launch a DDoS attack than to defend an infrastructure with the defending party. According to Arbor network, launching a DoS attack costs an average of \$66 per attack and can cause damage to the victim of around \$500 per minute ~\cite{ArborAttackSurvey}. Aura et. al~\cite{AURA2001DOS} proposed the first client puzzle authentication protocol for a DoS resilient system. Mankins et. al \cite{mankins2001mitigating} investigated methods for tuning the amount of resource consumption to access server resources based on client behavior, where the costs imposed can be either monetary or computational. In a similar vein, Wang and Reiter~\cite{wang:defending} investigate how clients can bid on puzzles through auctions. Ndibwile et. al~\cite{Ndibwile2015DoS} proposed web traffic authentication as a replacement for CAPTCHA-based defenses. Wu et. al~\cite{Wu2015Inflated} proposed a software puzzle framework that disqualifies the adversary's ability to gain an advantage by using a GPU to solve puzzles. A framework was put forth by Dean et. al~\cite{Dean2001TLS}  to reduce DoS in TLS servers. A DoS variant was introduced by Wood et. al~\cite{Wood2015DoSE}. Certain PoW defenses against DoS are layer-specific. The network layer of the proof-of-work system used by Parno et. al~\cite{Parno2007Portcullis} prioritizes users who use more CPU time to solve puzzles. The Heimdall architecture, which can detect any change in network flow in routers, was introduced by Chen et. al~\cite{Chen2010Bilateral}. When a change in network flow is identified for any new connection, a puzzle is generated and sent to the new user. The difficulty of the computational challenges used in the context of DoS attacks on the transport layer was recently assessed using game theory by Noureddine et. al~\cite{noureddinerevisiting}. Walfish et. al~\cite{walfish2006ddos} propose an alternative resource called communication capacity as a defense against application-layer flood attacks. Other research has concentrated on incorporating PoW puzzles into practical browsing experiences~\cite{le2012kapow, Chakraborty2022APD,CHAKRABORTY2022100335}.


\subsection{Automated DoS defense}
In this section, we revisit the literature on ensemble learning techniques for network traffic classification problems. Ensemble learning is a branch of supervised machine learning technique that aggregates the learning of multiple base learners to improve overall prediction accuracy~\cite{Polikar2006ensemble}. Like network traffic classification problems, each base learner is trained to become an expert in the local area of the total feature space. Gaikwad et. al~\cite{Gaikwad2015Bagging} proposed a bagging ensemble approach using REPTree base learners to improve classification over weaker AI models.  Gupta et. al~\cite{GUPTA2022Cost} suggested an IDS system that uses ensemble learning to address a class imbalance problem. The ensemble learner uses three base learners. First, the deep neural network classifies normal and suspicious traffic. Second, eXtreme Gradient Boosting is used to identify major attacks. Third, random forest is used to classify minor attacks. Zhou et. al~\cite{ZHOU2020Building} proposed feature selection process using ensemble learning in two stages. The first stage involves feature reduction using the heuristic method CFS and the Bat Algorithm (BA). The second stage involves aggregating C4.5 and Random Forest (RF) algorithms.  Jabbar et. al~\cite{Jabbar2017Cluster} suggested an ensemble classifier that uses Alternating Decision Tree (ADTree) and the k-Nearest Neighbor algorithm (kNN) as base AI models. Paulauskas and Auskalnis~\cite{Paulauskas2017Analysis} proposed an ensemble learner that employs four base classifiers: J48, C5.0, Naive Bayes, and Partial Decision List (PART) to improve classification results over individual AI models.


\section{Conclusion and Future Work}
\label{sec:conclusion}
In this paper, we design and evaluate \AlgoName a context-aware AI-assisted PoW framework that protects critical servers against DDoS. The underlying defensive strategy involves adaptively introducing latency on malicious users. To achieve this functionality, our framework employs an AI model that takes the context attributes from the incoming user request packet as input. The AI model computes deviation from normal activity patterns to output a context score. This score influences the difficulty level of a PoW puzzle that injects latency adaptively during communication. \AlgoName ensures that the ability of a malicious user to prolong the attack is constrained by adaptively introducing latency through PoW puzzles and compelling malicious users to expend more resources to complete an attack.

For future work, different design variants of \AlgoName can be configured to combat different DDoS attack types. PoW systems suffer from inherent pitfalls of resource wastage which can be circumvented by replacing the model with proof of stake (PoS) component. Additionally, alternate design can include enhanced human in loop strategy which provides control of the framework to the security personnel deploying the framework.





\bibliographystyle{plain} 
\bibliography{refs.bib}

\begin{thebibliography}{10}

\bibitem{ZHOU2020Building}
Building an efficient intrusion detection system based on feature selection and
  ensemble classifier.
\newblock {\em Computer Networks}, 174:107247, 2020.

\bibitem{GUPTA2022Cost}
Cse-ids: Using cost-sensitive deep learning and ensemble algorithms to handle
  class imbalance in network-based intrusion detection systems.
\newblock {\em Computers \& Security}, 112:102499, 2022.

\bibitem{OnlineCode}
Anonymous.
\newblock Github.
\newblock Online Website, 2022.
\newblock \url{https://github.com/voyagerinfinite/CAPoW}.

\bibitem{AURA2001DOS}
Tuomas Aura, Pekka Nikander, and Jussipekka Leiwo.
\newblock Dos resistant authentication with client puzzles.
\newblock In {\em Revised Papers from the 8th International Workshop on
  Security Protocols}, page 170–177, Berlin, Heidelberg, 2000.
  Springer-Verlag.

\bibitem{CHAKRABORTY2022100335}
Trisha Chakraborty, Shaswata Mitra, Sudip Mittal, and Maxwell Young.
\newblock Ai\_adaptive\_pow: An ai assisted proof of work (pow) framework for
  ddos defense.
\newblock {\em Software Impacts}, 13:100335, 2022.

\bibitem{Chakraborty2022APD}
Trisha Chakraborty, Shaswata Mitra, Sudip Mittal, and Maxwell Young.
\newblock A policy driven ai-assisted pow framework.
\newblock {\em 2022 52nd Annual IEEE/IFIP International Conference on
  Dependable Systems and Networks - Supplemental Volume (DSN-S)}, pages 37--38,
  2022.

\bibitem{Chen2010Bilateral}
Y.~{Chen}, W.~{Ku}, K.~{Sakai}, and C.~{DeCruze}.
\newblock A novel ddos attack defending framework with minimized bilateral
  damages.
\newblock In {\em 2010 7th IEEE Consumer Communications and Networking
  Conference}, pages 1--5, 2010.

\bibitem{Dean2001TLS}
Drew Dean and Adam Stubblefield.
\newblock Using client puzzles to protect {TLS}.
\newblock In {\em 10th USENIX Security Symposium (USENIX Security 01)},
  Washington, D.C., August 2001. USENIX Association.

\bibitem{Anind2001Context}
Anind~K. Dey.
\newblock Understanding and using context.
\newblock page 4–7, 2001.

\bibitem{Dwork1992Pricing}
Cynthia Dwork and Moni Naor.
\newblock Pricing via processing or combatting junk mail.
\newblock CRYPTO '92, page 139–147, Berlin, Heidelberg, 1992.
  Springer-Verlag.

\bibitem{Franklin1997Metering}
Matthew~K. Franklin and Dahlia Malkhi.
\newblock Auditable metering with lightweight security.
\newblock In {\em Proceedings of the First International Conference on
  Financial Cryptography}, FC '97, page 151–160, Berlin, Heidelberg, 1997.
  Springer-Verlag.

\bibitem{Gaikwad2015Bagging}
D.P. Gaikwad and Ravindra~C. Thool.
\newblock Intrusion detection system using bagging ensemble method of machine
  learning.
\newblock In {\em 2015 International Conference on Computing Communication
  Control and Automation}, pages 291--295, 2015.

\bibitem{Green2011Reconstructing}
Jeff Green, Joshua Juen, Omid Fatemieh, Ravinder Shankesi, Dong Jin, and
  Carl~A. Gunter.
\newblock Reconstructing hash reversal based proof of work schemes.
\newblock USA, 2011. USENIX Association.

\bibitem{gupta2020resource}
Diksha Gupta, Jared Saia, and Maxwell Young.
\newblock Resource burning for permissionless systems.
\newblock In {\em International Colloquium on Structural Information and
  Communication Complexity}, pages 19--44. Springer, 2020.

\bibitem{Jabbar2017Cluster}
M.~A. Jabbar, Rajanikanth Aluvalu, and S.~Sai~Satyanarayana Reddy.
\newblock Cluster based ensemble classification for intrusion detection system.
\newblock New York, NY, USA, 2017. Association for Computing Machinery.

\bibitem{Jakobsson1999Pudding}
Markus Jakobsson and Ari Juels.
\newblock Proofs of work and bread pudding protocols (extended abstract).
\newblock 1999.

\bibitem{juels:dos}
Ari Juels and John Brainard.
\newblock Client puzzles: A cryptographic countermeasure against connection
  depletion attacks.
\newblock In {\em Proceedings of the Network and Distributed System Security
  Symposium (NDSS)}, pages 151--165, 1999.

\bibitem{Laurie2004ProofofWorkPN}
Ben Laurie and Richard Clayton.
\newblock Proof-of-work" proves not to work.
\newblock 2004.

\bibitem{le2012kapow}
Tien Le, Akshay Dua, and Wu-chang Feng.
\newblock kapow plugins: protecting web applications using reputation-based
  proof-of-work.
\newblock In {\em 2nd Joint WICOW/AIRWeb Workshop on Web Quality}, pages
  60--63, 2012.

\bibitem{ArborAttackSurvey}
Vincent Lynch.
\newblock Everything you ever wanted to know about dos/ddos attacks, 2017.
\newblock
  \url{https://www.thesslstore.com/blog/everything-you-ever-wanted-to-know-about-dosddos-attacks/
  }.

\bibitem{mankins2001mitigating}
David Mankins, Rajesh Krishnan, Ceilyn Boyd, John Zao, and Michael Frentz.
\newblock Mitigating distributed denial of service attacks with dynamic
  resource pricing.
\newblock In {\em Proceedings of the Seventeenth Annual Computer Security
  Applications Conference}, pages 411--421. IEEE, 2001.

\bibitem{Ndibwile2015DoS}
Jema~David Ndibwile, A.~Govardhan, Kazuya Okada, and Youki Kadobayashi.
\newblock Web server protection against application layer ddos attacks using
  machine learning and traffic authentication.
\newblock In {\em 2015 IEEE 39th Annual Computer Software and Applications
  Conference}, volume~3, pages 261--267, 2015.

\bibitem{noureddinerevisiting}
M.~A. {Noureddine}, A.~M. {Fawaz}, A.~{Hsu}, C.~{Guldner}, S.~{Vijay},
  T.~{Ba\c{s}ar}, and W.~H. {Sanders}.
\newblock Revisiting client puzzles for state exhaustion attacks resilience.
\newblock In {\em Proceedings of the $49th$ Annual IEEE/IFIP International
  Conference on Dependable Systems and Networks (DSN)}, pages 617--629, 2019.

\bibitem{CICIDS2017}
University of~New~Brunswick.
\newblock Intrusion detection evaluation dataset (cic-ids2017).
\newblock Online Website, 2017.
\newblock \url{https://www.unb.ca/cic/datasets/ids-2017.html}.

\bibitem{DDoS-Saas}
Andrew Orlowski.
\newblock Meet ddosaas: Distributed denial of service-as-a-service.
\newblock Online Website, 2016.
\newblock
  \url{https://www.theregister.com/2016/09/12/denial_of_service_as_a_service/}.

\bibitem{Parno2007Portcullis}
Bryan Parno, Dan Wendlandt, Elaine Shi, Adrian Perrig, Bruce Maggs, and
  Yih-Chun Hu.
\newblock Portcullis: Protecting connection setup from denial-of-capability
  attacks.
\newblock SIGCOMM '07, page 289–300, New York, NY, USA, 2007. Association for
  Computing Machinery.

\bibitem{Paulauskas2017Analysis}
Nerijus Paulauskas and Juozas Auskalnis.
\newblock Analysis of data pre-processing influence on intrusion detection
  using nsl-kdd dataset.
\newblock In {\em 2017 Open Conference of Electrical, Electronic and
  Information Sciences (eStream)}, pages 1--5, 2017.

\bibitem{Polikar2006ensemble}
R.~Polikar.
\newblock Ensemble based systems in decision making.
\newblock {\em IEEE Circuits and Systems Magazine}, 6(3):21--45, 2006.

\bibitem{renjan2018dabr}
Arya Renjan, Karuna~Pande Joshi, Sandeep~Nair Narayanan, and Anupam Joshi.
\newblock Dabr: Dynamic attribute-based reputation scoring for malicious ip
  address detection.
\newblock In {\em 2018 IEEE International Conference on Intelligence and
  Security Informatics (ISI)}, pages 64--69. IEEE, 2018.

\bibitem{Rivest1996PayWordAM}
Ronald~L. Rivest and Adi Shamir.
\newblock Payword and micromint: Two simple micropayment schemes.
\newblock In {\em Security Protocols Workshop}, 1996.

\bibitem{Cisco2022Talos}
Cisco Talos.
\newblock Talos threat source, 2022.
\newblock \url{https://www.talosintelligence.com/}.

\bibitem{walfish2006ddos}
Michael Walfish, Mythili Vutukuru, Hari Balakrishnan, David Karger, and Scott
  Shenker.
\newblock {DDoS Defense by Offense}.
\newblock In {\em Proceedings of the 2006 Conference on Applications,
  Technologies, Architectures, and Protocols for Computer Communications
  (SIGCOMM)}, pages 303--314, 2006.

\bibitem{wang:defending}
XiaoFeng Wang and Michael~K. Reiter.
\newblock Defending against denial-of-service attacks with puzzle auctions.
\newblock In {\em Proceedings of the 2003 IEEE Symposium on Security and
  Privacy}, pages 78--92, 2003.

\bibitem{waters:new}
Brent Waters, Ari Juels, Alex Halderman, and Edward Felten.
\newblock New client puzzle outsourcing techniques for {DoS} resistance.
\newblock In {\em Proceedings of the 11th ACM Conference on Computer and
  Communications Security (CCS)}, pages 246--256, 2004.

\bibitem{Wood2015DoSE}
Paul Wood, Christopher Gutierrez, and Saurabh Bagchi.
\newblock Denial of service elusion (dose): Keeping clients connected for less.
\newblock In {\em 2015 IEEE 34th Symposium on Reliable Distributed Systems
  (SRDS)}, pages 94--103, 2015.

\bibitem{Wu2015Inflated}
Yongdong Wu, Zhigang Zhao, Feng Bao, and Robert~H. Deng.
\newblock Software puzzle: A countermeasure to resource-inflated
  denial-of-service attacks.
\newblock {\em IEEE Transactions on Information Forensics and Security},
  10(1):168--177, 2015.

\bibitem{Ozdel2022Payload}
Süleyman Özdel, Pelin Damla~Ateş, Çağatay Ateş, Mutlu Koca, and Emin
  Anarım.
\newblock Network anomaly detection with payload-based analysis.
\newblock In {\em 2022 30th Signal Processing and Communications Applications
  Conference (SIU)}, pages 1--4, 2022.

\end{thebibliography}

\end{document}